\pdfoutput=1

\def\lj{{\lambda_J}}
\def\bSG{{\beta_{_{SG}}}}
\def\gf{{\gamma^{^{L=1}}_{_{MAX}}}}
\def\gs{{\gamma^{^{L=10}}_{_{MAX}}}}

\documentclass[aps,prb,twocolumn,showpacs,superscriptaddress,groupaddress,amsmath,amssymb,floatfix]{revtex4}
\usepackage{epsfig,graphics}
\usepackage{multirow}
\usepackage {amsmath, amssymb,wasysym}
\usepackage{color}
\usepackage{dcolumn}
\usepackage{bm}
\usepackage{textcomp}
\usepackage[T1]{fontenc}

\begin{document}

\title{\bf Switching times in long-overlap Josephson junctions subject to thermal fluctuations and non-Gaussian noise sources}

\author{Davide Valenti\footnote{e-mail address: davide.valenti@unipa.it}}
\affiliation{Dipartimento di Fisica e Chimica, Group of Interdisciplinary Theoretical Physics, Universit\`{a} di Palermo and CNISM,
Unit\`{a} di Palermo, Viale delle Scienze, Edificio 18, I-90128 Palermo, Italy}
\author{Claudio Guarcello\footnote{e-mail address: claudio.guarcello@unipa.it}}
\affiliation{Dipartimento di Fisica e Chimica, Group of Interdisciplinary Theoretical Physics, Universit\`{a} di Palermo and CNISM, 
Unit\`{a} di Palermo, Viale delle Scienze, Edificio 18, I-90128 Palermo, Italy}
\author{Bernardo Spagnolo\footnote{e-mail address: bernardo.spagnolo@unipa.it}}
\affiliation{Dipartimento di Fisica e Chimica, Group of Interdisciplinary Theoretical Physics,
Universit\`{a} di Palermo and CNISM, Unit\`{a} di Palermo,
Viale delle Scienze, Edificio 18, I-90128 Palermo, Italy}
\affiliation{Istituto Nazionale di Fisica Nucleare, Sezione di Catania,\\
Via S. Sofia 64, I-90123 Catania, Italy}

\begin{abstract}
We investigate the superconducting lifetime of long current-biased Josephson junctions, in the presence 
of Gaussian and non-Gaussian noise sources. In particular, we analyze the dynamics of a Josephson junction as a
function of the noise signal intensity, for different values of the parameters of the system and external driving currents. 
We find that the mean lifetime of the superconductive state is characterized by nonmonotonic behavior as a function of 
noise intensity, driving frequency and junction length. We observe that these nonmonotonic behaviours are connected 
with the dynamics of the junction phase string during the switching towards the resistive state. An important role is played 
by the formation and propagation of solitons, with two different dynamical regimes characterizing the dynamics of the 
phase string. Our analysis allows to evidence the effects of different bias current densities, that is a simple spatially 
homogeneous distribution and a more realistic inhomogeneous distribution with high current values at the junction edges. 
Stochastic resonant activation, noise enhanced stability and temporary trapping phenomena are observed in the system investigated. 
\end{abstract}

\date{\today}

\maketitle

\section{Introduction}
\label{Introd}

During last decades the interest in superconductor physics and its applications had a remarkable development. In this context an
important role is played by improvements made in devising and realizing Josephson junction (JJ) based devices. 
In fact, great attention has been paid to JJs as superconducting quantum bits~\cite{Wen07,Kim06,Zor06,Ber03}, nanoscale 
superconducting quantum interference devices for detecting weak flux changes~\cite{Wu08,Lev13}, and threshold noise 
detectors~\cite{Gra08,Urb09,Fil10,Add12}. Moreover JJs are typical out of equilibrium systems characterized by tilted or
switching periodic potentials~\cite{Rei01,Dub05}.

The behavior of these systems is strongly influenced by environmental perturbations, and specifically by the presence of noise 
source responsible for decoherence phenomena~\cite{Kim06,Xu05}. The role played by random fluctuations in the dynamics 
of these devices has recently solicited a large amount of work and investigation on the effects both of thermal and non-thermal 
noise sources on the transient dynamics of Josephson junctions~\cite{Hua07,Nov09,LeM09,Bil08,Pel07,Tob04}. 
The noise current signal is caused by the stochastic motion of the charge carriers, namely the Cooper pairs in a superconductor.  
While thermal noise is originated by the thermal motion of the charge carriers, non-thermal noise signal are related to their 
scattering and transmission. Non-Gaussian noise appears when the conductor, or the superconductor, is in a non-equilibrium 
state because of the presence of a bias voltage or current. In the last decade, theoretical progress allowed one to calculate the 
entire probability distribution of the noise signal and its cumulants, performing a \textit{full counting statistics} 
of the current fluctuations~\cite{Nov09}. Moreover, the presence of non-Gaussian noise signals has been found experimentally 
in many systems~\cite{Hua07,Pel07,Mon84,Shle95,Dyb04,Sou08}. As an example in a wireless ad hoc network with a Poisson 
field of co-channel users, the noise has been well modeled by an $\alpha$-stable distribution~\cite{Sou08}. Non-equilibrated 
heat reservoir can be considered as a source of non-Gaussian noise  sources~\cite{Mon84,Shle95,Dyb04}. 
Specifically, the effect of non-Gaussian noise on the average escape time from the superconducting metastable 
state of a current biased JJ, coupled with non equilibrium current fluctuations, has been experimentally 
investigated~\cite{Hua07,Pel07}.

Recently, the characterization of JJs as detectors, based on the statistics of the escape times, has been 
proposed~\cite{Gra08,Urb09,Fil10,Add12,Ank07,Suk07,Kop13}. Specifically, the statical analysis of the switching from the metastable 
superconducting state to the resistive running state of the JJ has been proposed to detect weak periodic signals embedded in a 
noise environment~\cite{Fil10,Add12}. Moreover, the rate of escape from one of the metastable wells of the tilted washboard 
potential of a JJ encodes information about the non-Gaussian noise present in the input signal~\cite{Gra08,Urb09,Ank07,Suk07,Kop13}.

\begin{figure*}[htbp!!]
\centering
\includegraphics[width=171mm]{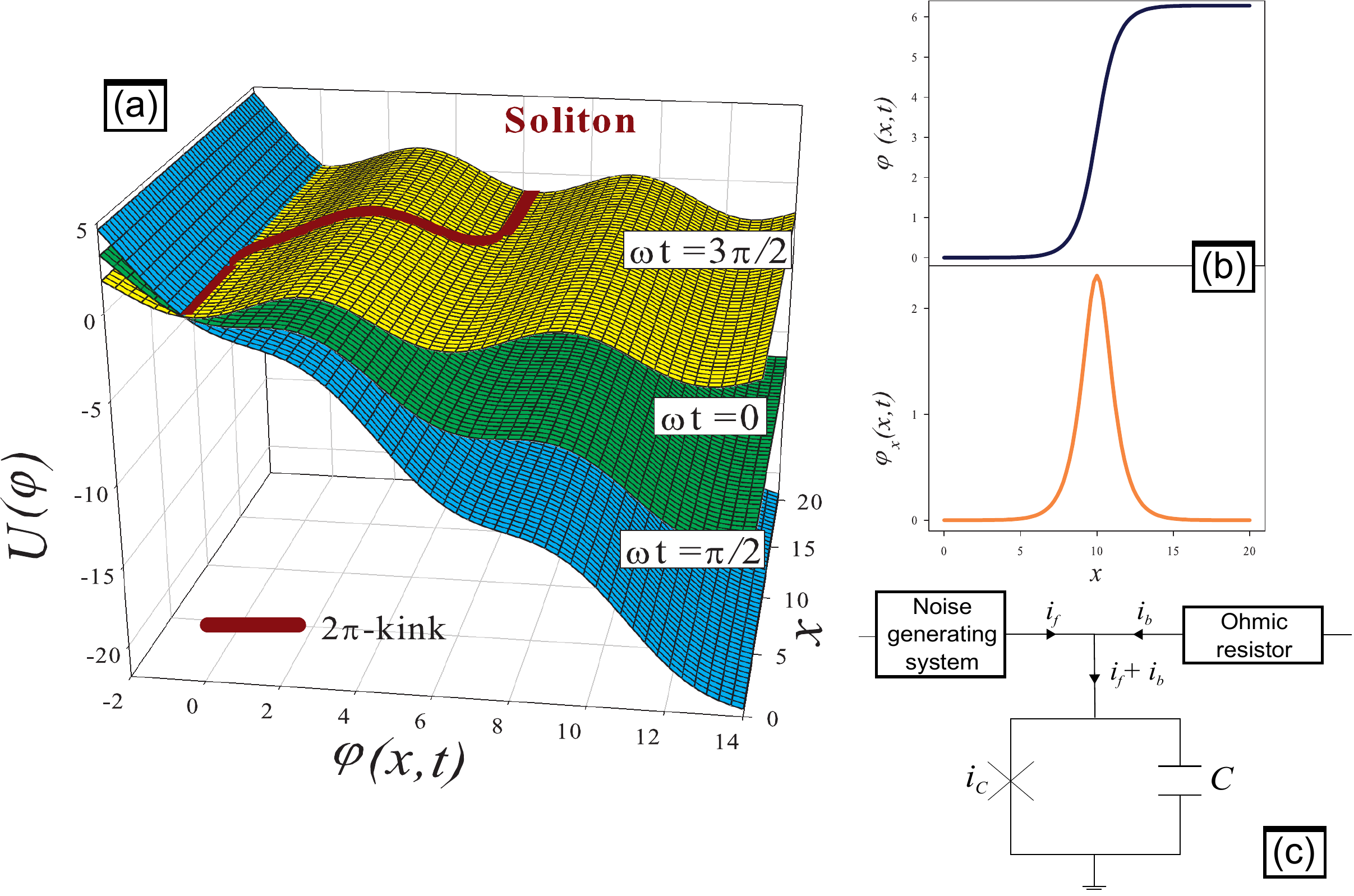}
\qquad\qquad \caption{(Color online) (a) Washboard potential at 3 different times with a soliton wave ($2\pi$-kink) on the
highest profile; (b) Soliton (Eq.~(\ref{SGkink})) and corresponding fluxon profile (Eq.~(\ref{phiX-H})); (c) Circuit diagram of
a JJ noise detector: a JJ with critical current $i_C$ is biased in a twofold way.}
\label{fig2+3}
\end{figure*}

Motivated by these studies and the importance of the problem of the transient dynamics of a JJ interacting with a 
noisy environment, we try to understand how non-Gaussian noise sources affect the switching times in long JJs.
In light of this, our work is devoted to investigate the response of a superconductive device to the solicitations of 
both deterministic and stochastic external perturbations, due to thermal fluctuations~\cite{Gord06,Gord08,Pank04} or 
connected with the variability of bias current and magnetic field~\cite{Hua07,Pel07}. In particular, we analyze 
the system dynamics, modelling environmental random fluctuations by noise sources with different, 
Gaussian and non-Gaussian, statistical distributions. The superconducting device is a long Josephson 
junction (LJJ), that is a device in which one dimension is much longer than the Josephson penetration depth 
$\lj$ of the junction. The JJs considered in our study are arranged in the \textit{overlap} geometrical configuration. 
These devices can work in two different conditions: i) superconducting regime, that corresponds to the localization 
of the order parameter, that is the phase difference across the junction, in a metastable state of the washboard potential; 
ii) resistive regime with a dissipative voltage-current relation, corresponding to an escape event of the phase difference 
from the metastable state (see Fig.~\ref{fig2+3}a). The superconductive phase is subject to both thermal and non-Gaussian noise, 
due to an external driving force. We note that, effects of Gaussian~\cite{Gord06,Gord08,Pank04,Auge08,Gord2_08} and 
non-Gaussian~\cite{Gra08,Urb09,Ank07,Suk07,Auge10} noise sources on short JJs have been thoroughly studied, while 
analyses of the phase dynamics of long JJs have been performed only in the presence of thermal 
fluctuations~\cite{Auge09,Fedo07,Fedo08,Fedo09}. Moreover, noise induced effects due to thermal fluctuations, such as 
\textit{resonant activation} (RA), or \textit{stochastic resonant activation}~\cite{Doe92,Man00}, and 
\textit{noise enhanced stability}~\cite{Man96,Agu01} (NES), have been theoretically predicted in 
overdamped JJs~\cite{Gord06,Gord08,Pank04,Auge08,Gord2_08,Auge10}, and experimentally in underdamped 
JJs~\cite{Yu03,Sun07,Pan09}. It is worthwhile to note that experimental works on the realization of overdamped JJ with 
non-hysteretic current voltage and high temperature stability have been performed~\cite{Feb10}.

After the seminal paper of Tobiska and Nazarov~\cite{Tob04}, Josephson junctions used as threshold detectors allow to 
study non-Gaussian features of current noise~\cite{Ank07,Suk07}. Specifically, when a JJ leaves the metastable zero voltage 
state it switches to a running resistive state and a voltage appears across the junction. Therefore, it is possible to measure directly 
in experiments the escape times or switching times and to determine its probability distribution~\cite{Yu03,Sun07,Pan09,Dev84,Dev85,Cas96}. 
A typical simplified realization of a JJ noise detector is shown in Fig.~\ref{fig2+3}c. The fluctuating current $i_f$, produced by 
the noise generating system, is added to the bias current $i_b$ and drives the JJ, characterized by a critical current $i_C$ 
and a capacitance $C$. The switching times of the junction can be directly measured using the time-domain 
technique~\cite{Yu03,Sun07,Pan09,Han01,Yu02}. For each switching event the bias current is ramped up to a value $i_b$, which 
is very close to the critical current $i_C$ and it is maintained constant for a period of waiting time. To record the switching time, 
the voltage across the junction is sent to a timer-counter, which is triggered by the sudden jump from zero voltage state to finite voltage state. 
The bias current is then decreased to zero, the junction returns to the zero voltage state, and a new cycle starts again. For JJs 
working in over damped regime, the superconducting state is restored automatically, without necessity to decrease the bias current to zero. 
The process is repeated enough times to obtain a statistically significant ensemble of switching times (ST). 
Typical frequency range of a detector of non-Gaussian noise, based on a long JJ working in an overdamped regime, as in our investigation, 
is from $10$ GHz to $600$ GHz. Of course, higher frequencies can be obtained with a long JJ in overlap geometry, but the experimental 
setup should be more complicated and very expensive. Concerning the physical range of feasibility of the other main parameters of 
the junction, typical values are: JJ length $L$ from $0.1 \lambda_J$ to $20 \lambda_J$, with the Josephson penetration depth 
$\lambda_J$ in the range $[10,20]~\mu$m; range of the critical current $[5,15]$ mA.

In this paper we investigate how the simultaneous action of an external oscillating driving force and a fluctuating signal 
affects the permanence time inside the metastable state of a LJJ. In particular, we concentrate on the escape time, that is the time 
the junction takes to switch from the superconducting state to the resistive regime, calculating the \emph{mean switching time} (MST) 
obtained by averaging on a sufficiently large number of numerical realizations. The analysis is performed varying also 
the frequency of the driving current, the length of the junction, and the amplitude of the noise signal modeled by using 
different $\alpha$-stable (or \textit{L\'evy}) distributions. These statistics allow to describe real situations~\cite{Szaj01} 
in which the evolution shows abrupt jumps and very rapid variations of parameters, called \emph{L\'evy flights}. 
L\'evy-type statistics is observed in various scientific areas, where scale-invariance phenomena take 
place~\cite{Che06,Met00,Uch03,Dub09}. For a recent short review on L\'evy flights see Dubkov \textit{et al.}~\cite{Dub08}, 
and references there. Applications and other research fields in which observed evolutions are well reproduced using 
L\'evy statistics are quite numerous, ranging from biology~\cite{West94}, zoology~\cite{Reyn94,Sims08,Reyn08}, social 
systems~\cite{Broc06} and financial markets~\cite{Mant95}, to geological~\cite{Shle71} and atmospheric data~\cite{Ditl99}.

The dynamics of the phase difference of the LJJ, analyzed within the sine-Gordon (SG) formalism~\cite{Fedo07,Fedo08,Baro82,Likh86}, 
is characterized by the formation and propagation of particular wave packets, called \emph{solitons}~\cite{Usti98,Butt81}. 
Their presence is strongly connected with the penetration of the magnetic flux quanta, i.e. \emph{fluxons}~\cite{McLa78,Dueh83} 
(the magnetic soliton), travelling through the junction during the switching towards the resistive state (see Fig.~\ref{fig2+3}b). Here we recall that several 
systems governed by SG equation show evidence of soliton motion, including not only JJs~\cite{Abdu05,Pfei06,Fedo11,Pank12,GaNu12,Kim12,Gule12} 
but also the relativistic field theory, mechanical transmission lines, and atomic, particle and condensed matter physics. 
A peculiar dynamics is also present in the superconducting device analyzed in this work.

Finally, it is worth nothing that for low phase values, $sin(\varphi)\approx \varphi$, the SG equation approaches the Klein-Gordon 
one~\cite{Wic55}. Nevertheless, the exact solutions are known only for the simplest \emph{unperturbed} SG differential equation, in the absence of damping, 
driving and fluctuating terms~\cite{Baro82}.

The paper is organized as follows. In the next section the sine-Gordon model is presented. In Sec. III we briefly review the statistical 
properties of the L\'evy noise, showing some peculiarities
of different $\alpha$-stable distributions. Section IV gives computational details. In Sec. V the theoretical results for the behaviors of the MST 
as a function of the junction length, frequency of the external driving current and noise intensity with homogeneous and inhomogeneous bias 
current, are shown and analyzed. This analysis has been carried out at very low temperatures 
of the system, around the \textit{crossover}
temperature. 

Below this temperature, the phase difference over the junction behaves quantum mechanically, the escape events occur primarily 
by quantum tunneling through the barrier, and the thermal fluctuations can be neglected. Therefore, only the effects of non-Gaussian 
noise have been analyzed. The transient dynamics of a long JJ subject to thermal fluctuations and non-Gaussian, L\'evy type, 
noise sources is investigated in Sec. VI. Finally, in Sec. VII we draw the conclusions.

\section{The SG Model}
\label{SG Model}

The electrodynamics of a normal JJ is described by a nonlinear partial differential equation for the order parameter $\varphi$, that is the 
sine-Gordon equation~\cite{Baro82,Likh86}. 
Here $\varphi$ is the phase difference between the wave functions describing the superconducting condensate in the two electrodes. 
Our analysis includes a quasiparticle tunneling term and an additional stochastic contribution, $i_{f}(x,t)$, representing the noise effects. 
However, the surface resistance of the superconductors is neglected. The resulting \emph{perturbed} SG equation reads

\vskip-0.2cm
\begin{eqnarray}
\bSG \varphi_{tt}(x,t)+\varphi_{t}(x,t)&-&\varphi_{xx}(x,t) \nonumber \\
= i_b(x,t)&-&\sin(\varphi(x,t))+i_{f}(x,t),
\label{sine_gordon}
\end{eqnarray}
where a simplified notation has been used, with the subscript indicating the partial derivative of $\varphi$ in that variable. This notation will be 
used throughout all paper. In Eq.~(\ref{sine_gordon}), the fluctuating current density $i_{f}(x,t)$ is the sum of two contributions, 
a Gaussian thermal noise $i_T(x,t)$ and an external non-Gaussian noise source $i_{nG}(x,t)$
\begin{equation}
i_{f}(x,t) = i_T(x,t) + i_{nG}(x,t) .
\label{fluct.current}
\end{equation}

The SG equation is written in terms of the dimensionless $x$ and $t$ variables, that are the space and time coordinates normalized 
respectively to the Josephson penetration depth $\lj$ and to the inverse of the characteristic frequency $\omega_J$ of the junction. 
Moreover, $\bSG=\omega_J RC$, where $R$ and $C$ are the effective normal resistance and capacitance of the junction. 
The terms $i_b(x,t)$ and $sin(\varphi)$ of Eq.~(\ref{sine_gordon}) are respectively the bias current and supercurrent, both 
normalized to the JJ critical current $i_C$. Eq.~(\ref{sine_gordon}) is solved imposing the following boundary conditions
\begin{equation}
\varphi_{x}(0,t) = \varphi_{x}(L,t) = \Gamma, 
\label{boundary}
\end{equation}
where $\Gamma$ is the normalized external magnetic field. Hereinafter we impose $\Gamma = 0$.

The two-dimensional time-dependent tilted potential, named \emph{washboard potential}, is given by
\begin{equation}
U ( \varphi ,x,t )=1-\cos(\varphi ( x,t ))-i_b( x,t ) \thinspace \varphi ( x,t ), 
\label{Washboard}
\end{equation}
and shown in Fig.~\ref{fig2+3}a. In the same figure is shown a phase string in the potential profile~(\ref{Washboard}), along which 
it moves during the switching dynamics. Specifically, the washboard potential is composed by a periodical sequence of 
peaks and valleys, with minima and maxima satisfying the following conditions
\begin{eqnarray}
\label{min_max}
\varphi_{min}&=&\arcsin( i(x,t))+2n\pi  \nonumber \\
\varphi_{max}&=&(\pi-\arcsin( i(x,t)))+2n\pi
\end{eqnarray}
with $n=0,\pm1,\pm2,...$ .

The bias current is given by
\begin{equation}
i_b(x,t) = i_b(x)+A \thinspace \sin(\omega t),
 \label{DrivingCurrent}
\end{equation}
where $A$ and $\omega$ are amplitude and frequency (normalized to $\omega_J$) of the dimensionless driving current. This time
dependence is normalized to the inverse of the JJ characteristic frequency $\omega_J$.

\begin{figure}[htbp!!]
\centering
\includegraphics[width=90mm]{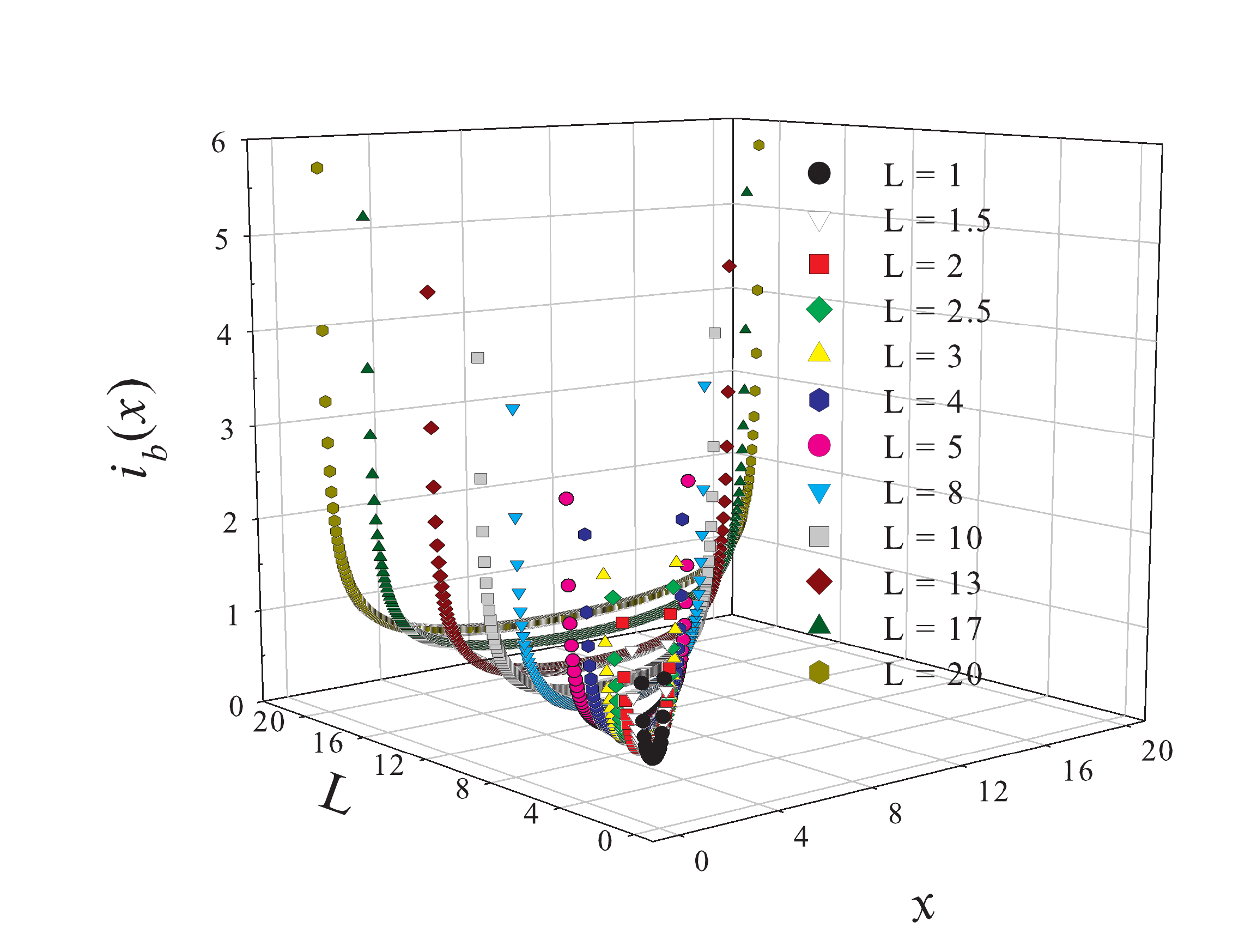}
\caption{(Color online) Inhomogeneous bias current density~(see Eq.~(\ref{BiasCurrent})) along JJs, for $i_0 = 0.9$ and different
values of junction length.} 
\label{fig1}
\end{figure}
The $i_b(x)$ term is a dimensionless current that, in the phase string picture, represents the initial slope of the potential profile. 
Different regimes of spatial dependence can be considered, obtaining in particular the two following current distributions~\cite{Samu85}
\begin{eqnarray}
i_b(x)=\left\{\begin{matrix}
i_0 \hspace{16 mm} homogeneous \\
\\
\frac{i_0 \thinspace L}{\pi \sqrt{x \thinspace (L-x)}} \hspace{8 mm} inhomogeneous.
\end{matrix}\right.    
\label{BiasCurrent}
\end{eqnarray}
The more realistic inhomogeneous condition provides strong current contributions at the edges of the junction. This is shown in
Fig.~\ref{fig1}, for $i_0=0.9$ and $L$ ranging between $1$ and $20$. In these conditions, the phase of the cells in the edges of
the junction can flow along the potential without resistance, so that the soliton formation occurs mostly in these parts of the junction.

The unperturbed SG equation, in the absence of damping, bias and noise, is given by
\begin{equation}
\varphi_{xx}(x,t)-\varphi_{tt}(x,t)=\sin(\varphi(x,t)).
\label{SGunperturbed}
\end{equation}
This equation admits solutions in the traveling wave form $f=\varphi(x-ut)$~\cite{Baro82}
\begin{equation}
\varphi(x-ut)=4\arctan \left \{ \exp \left [ \pm \frac{\left(x-ut \right )}{\sqrt{1-u^2}} \right ] \right \}, 
\label{SGkink}
\end{equation}
where $u$ is the wave propagation velocity normalized to the speed of light, and is called \emph{Swihart velocity}. Eq.~(\ref{SGkink})
represents a single \emph{kink}, or \emph{soliton}, that is a $2\pi$ variation in the phase values. The signs $+$ and $-$ indicate the
two opposite directions of propagation, corresponding to $2\pi$-kink (soliton) and $2\pi$-antikink (antisoliton), respectively. In this
framework, $\varphi$ gives a normalized measure of the magnetic flux through the junction, so that Eq.~(\ref{SGunperturbed}) can also
represent the motion of a single fluxon (or antifluxon). In fact, starting from simple electrodynamic considerations~\cite{Baro82}, it
is possible to obtain a simple relation between the magnetic field $H(y)$ and the spatial derivative of the phase difference
\begin{equation}
\varphi_x=\frac{2e}{\hbar c}d H(y), \label{phiX-H}
\end{equation}
where $d=\lambda_L+\lambda_R+t$ is the magnetic penetration, $\lambda_L$ and $\lambda_R$ are the London depths in the left and
right superconductors and $t$ is the interlayer thickness. In our LJJ model, if the junction is extended along $x$ and short along
$z$, the magnetic field points along $y$, so that $H(y)\equiv H$. Integrating Eq.~(\ref{phiX-H}) over the entire JJ length the
following relation is obtained
\begin{equation}
\varphi(L)-\varphi(0)=\frac{2e}{\hbar c}\Phi_H=2\pi\frac{\Phi_H}{\Phi_0}, \label{phi-H}
\end{equation}
where $\Phi_H$ is the magnetic flux through the junction and $\Phi_0=hc/2e$ is the fluxon. If the phase string has a portion
lying in the first valley and a portion inside the $n-$valley, from Eq.~(\ref{min_max}), the phase difference is equal to $2\pi n$.
Therefore the magnetic flux will be equal to
\begin{equation}
2\pi n = 2\pi\frac{\Phi_H}{\Phi_0} \hspace{15 mm}  \Phi_H = n \Phi_0.   \label{FluxH}
\end{equation}
If the phase evolution shows a single $2\pi$-kink, a single fluxon will propagate along the junction, as shown in Fig.~\ref{fig2+3}b.
Here the washboard potential is represented at three different times $ t=0, \frac{\pi}{2\omega}, \frac{3\pi}{2\omega}$, corresponding to zero initial slope,
maximum and minimum slope, respectively. The line on the highest potential profile represents a soliton between two adjacent valleys. 
The panel (b) of the same figure shows a soliton and the shape of the correspondent fluxon, that is the values of the $x$-derivative 
of $\varphi$, along the junction length in a generic time $t\rq{}$.

\section{The L\'evy Statistics}
\label{Levy_noise}
\vskip-0.2cm 

In order to motivate the use of $\alpha$-stable (or L\'evy) distributions we recall some cases~\cite{Woyc01} in which non-Gaussian stable 
statistics is used to model experimental data with asymmetric and heavy tailed distributions, closely linked with the 
Generalized Central Limit Theorem~\cite{Bert96,Sato99,Gned54,deFi75,Khin36,Khin38,Fell71}. Here we briefly review the concept of 
stable distribution. A random non-degenerate variable is stable if
\begin{eqnarray}
\forall n\in\mathbb{N}, \exists (a_n,b_n) \in  \mathbb{R}^+\times\mathbb{R}: \nonumber \\
X+b_n=a_n\sum_{j=1}^{n} X_j,   \label{AlfaStable}
\end{eqnarray}
where the $X_j$ terms are independent copies of $X$. Moreover $X$ is strictly stable if and only if $b_n=0 \,\,\, \forall n$. The well
known Gaussian distribution stays in this class. This definition does not provide a parametric handling form of the stable
distributions. The characteristic function, however, allows to deals with them. The general definition of characteristic function for a
random variable $X$ with an associated distribution function $F(x)$ is
\begin{equation}
\phi (u) = \left < e^{iuX} \right > = \int_{-\infty}^{+\infty}e^{iuX}dF(x).
\label{GenerealCharFunc}
\end{equation}
Following this statement, a random variable $X$ is said stable if and only if
\begin{eqnarray}
\exists (\alpha ,\sigma, \beta, \mu )&\in&]0, 2]\times \mathbb{R}^+\times [-1, 1]\times \mathbb{R}: \nonumber \\
X&\overset{d}{=}&\sigma Z+\mu,    
\label{XStableFunc}
\end{eqnarray}
where $Z$ is a random number. Accordingly one obtains
\begin{eqnarray}
\phi(u)=\left\{\begin{matrix}
\exp \left \{-\left | u \right |^\alpha\left [ 1-i\beta \tan\frac{\pi\alpha}{2}(\textup{sign}u) \right ]\right \} \,\,\,\, \alpha\neq 1\\
\\
\exp \left \{-\left | u \right |\left [ 1+i\beta\frac{2}{\pi}(\textup{sign}u)\log \left | u \right |\right ]\right \} \,\,\,\, \alpha=1
\end{matrix}\right.    \label{XCharFunc}
\end{eqnarray}
\begin{table}
\begin{center}
\begin{tabular}{l|c|cc|c|}
Distribution&Abbr.&P($x$)&&$S_{\alpha}(\sigma, \beta, \mu)$\\ \hline
Gaussian&(G)&\Large$\frac1{\sqrt{2\pi}\sigma}{e^{-\frac{(x-\mu)^2}{2\sigma^2}}}$\normalsize&$x\in\mathbb{R}$&$S_2(\sigma, 0, \mu)$\\ \hline
Cauchy-Lorentz &(CL)&\Large$\frac{\sigma/\pi }{\sigma ^2+(x-\mu)^2}$\normalsize &$x\in\mathbb{R}$&$S_1(\sigma, 0, \mu)$\\ \hline
L\'evy-Smirnov &(LS)&\Large$\sqrt{\frac{\sigma}{2\pi}}\frac{e^{-\frac{\sigma}{2(x-\mu)}}}{(x-\mu)^{3/2}}$\normalsize&$x\geq\mu$&$S_{\frac12}(\sigma, 1, \mu)$\\ 
\hline
\end{tabular}
\end{center}
\caption{Closed form of the stable distributions and characteristic values of parameters.}\label{LevyTable}
\end{table}
in which
\begin{eqnarray}
\textup{sign}u=\left\{\begin{matrix}
\pm1\,\,\, &u&\gtrless0\\
0 &u&=0
\end{matrix}\right.   \label{signU}
\end{eqnarray}
represents the $\textup{sign}$ function.

This definition of $X$ requires four parameters: a \emph{stability index} (or characteristic exponent) $\alpha\in]0,2]$, 
an \emph{asymmetry parameter} $\beta$ with $\left| \beta \right |\leq 1$ and two real numbers $\sigma>0$  and
$\mu$ that determine the outward of the distribution and are called, for this reason, \emph{shape parameters}. 
The names of these two parameters indicate their physical meaning. Specifically $\beta=0$ ($\beta\neq0$) gives a 
symmetric (asymmetric) distribution, while $\alpha$ determines how the tails of distribution go to zero. 
For $\alpha<2$ the asymptotic behaviour is characterized by a power law, while $\alpha=2$ and $\beta=0$ give a 
Gaussian distribution. The stable distribution, obtained setting $\sigma=1$ and $\mu=0$, is called \emph{standard}. 
We denote every $\alpha$-stable distribution with the symbol $S_{\alpha}(\sigma, \beta, \mu)$. 
Only a few number of L\'evy distributions has a probability density function known in explicit form, as shown in 
Table~(\ref{LevyTable}). Here the abbreviations for some peculiar distributions, used in the rest of this work, are also listed.
\begin{figure}[htbp!!]
\centering
\includegraphics[width=87mm]{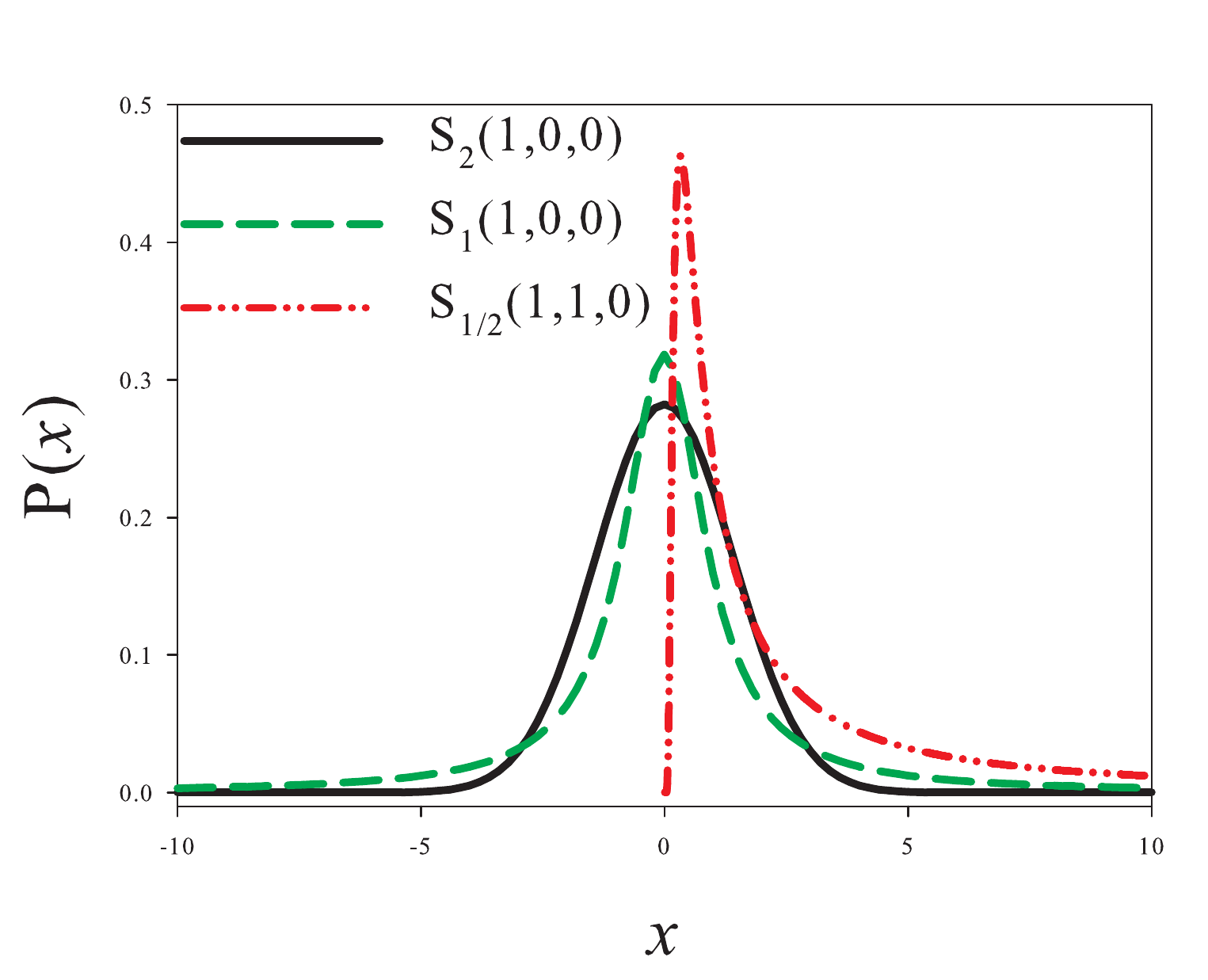}
\qquad\qquad \caption{(Color online) Probability density functions for Gaussian (G) (solid line), Cauchy-Lorentz (CL) (dashed line) 
and L\'evy-Smirnov (LS) (dashed-dotted line) distributions.}
\label{fig4}
\end{figure}
The G (Gaussian) and CL (Cauchy-Lorentz) distributions (both with $\beta=0$) are symmetrycal with respect to $x=0$, while 
the LS (L\'evy-Smirnov) distributions (normal and reflected) are skewed to the right ($\beta=+1$) or left ($\beta=-1$) side. 
The three distributions of Table~(\ref{LevyTable}) are plotted in Fig.~\ref{fig4}. The reflected (with respect to the vertical axis) 
LS distribution, obtained setting $\beta=-1$, is not shown. The asymmetrical structure of the LS distribution 
is evident, with a heavy tail and a narrow peak located at a positive value of $x$. The CL distribution, in comparison 
with the Gaussian one, presents tails much higher and a central part of the distribution more concentrated around the 
mean value. For short times, the values extracted from a CL distribution determine trajectories characterized 
by \emph{limited space displacement}: this can be explained noting that the CL statistics is characterized, around the mean, 
by a narrower form respect to  the Gaussian one. For longer times, however, heavy tails cause the occurrence of events 
with large values of $x$, whose probability densities are non-neglectable. The use of CL and LS statistics allows to 
consider rare events, corresponding to large values of $x$, because of the fat tails of these distributions. These events 
correspond to the L\'evy flights previously discussed. The algorithm used in this work to simulate L\'evy 
noise sources is that proposed by Weron~\cite{Wero96} for the implementation of the Chambers method~\cite{Cham76}.

\section{Computational Details}

\label{CompDetail}
\vskip-0.2cm 

We study the JJ dynamics in the SG overdamped regime, setting $\bSG=0.01$. The time and spatial steps are fixed at 
$\Delta t=0.05$ and $\Delta x=0.05$. In order to obtain the mean values we perform a suitable number ($N=5000$) of 
numerical realizations. Throughout the whole paper we use the words \emph{string}, referring to the entire junction, and 
\emph{cell} to indicate each of the elements with dimension $\Delta x$, which compose the junction. The washboard 
potential valley labeled with $n=0$ (Eq.~(\ref{min_max})) is chosen as initial condition for solving Eq.~(\ref{sine_gordon}), i.e. 
$\varphi_0=\arcsin(i_b(x,0))=\arcsin( i_b(x))$. In our model there are no barriers, neither absorbing nor reflecting, surrounding the 
initial metastable state, and the value of MST calculated is the nonlinear relaxation time (NLRT)~\cite{Dub04}. After a first exit, 
other temporary trapping events are permitted: during the time evolution each cell can return into the initial potential well, 
contributing again to the final value of MST, indicate as $\tau$. This agrees with the definition, proposed by Malakhov~\cite{Mala02}, 
for the mean permanence time of the phase $\varphi$ inside the interval [$-\pi,\pi$]
\begin{equation}
\tau=\int_{0}^{\infty }tw(t)dt=\int_{0}^{\infty}P(t)dt ,
 \label{tauNLRT}
\end{equation}
where $P(t)$ is the probability that $\varphi\in[-\pi,\pi]$ and $w(t)=\partial P(t)/\partial t$. For each cell and for each realization 
the numerical calculation of $P(t)$ is performed by considering 
\begin{eqnarray}
P_{ij}(t) = \left\{\begin{matrix}
 1 \iff \varphi\in[-\pi,\pi]\\
\\
0 \iff \varphi\notin [-\pi,\pi] ,
\end{matrix}\right.
 \label{P_t}
\end{eqnarray}
where $P_{ij}$ is the probability that in the i-th realization for the j-th cell $\varphi\in[-\pi,\pi]$. Summing $P_{ij}(t)$ over the 
total number $N_{cells}$ of string elements, and averaging first over the total number of cells, then 
over the total number $N$ of realizations, we find the probability that the entire string is in the superconducting state at time $t$
\begin{equation}
P(t) = \frac{1}{N \thinspace N_{cells}}\sum_{i=1}^{N}\sum_{j=1}^{N_{cells}}P_{ij}(t)
\label{P_averaged}
\end{equation}
The maximum time value used to perform the integral of Eq.~(\ref{tauNLRT}) has to be set large enough so that temporary trapping 
events, in the metastable state, can occur. We replace therefore the upper limit of the integral, $\infty$, with a maximum time 
$t_{MAX}=100$, obtaining the \emph{mean switching time}
\begin{equation}
\tau = \int_{0}^{t_{MAX}} P(t)dt.
 \label{tauNLRT_bis}
\end{equation}
The whole procedure is repeated for the three noise statistics analyzed in the previous section, obtaining the 
behaviour of the MST $\tau$ in the presence of different sources of L\'{e}vy noise.

\section{Effects of non-Gaussian noise} 
\label{Non-Gauss}

The analysis is carried out looking at the MST variations as a function of the junction length $L$, noise intensity $\gamma$ 
and frequency $\omega$ of the driving signal. The $i_0$ values choosen are $0.5$ and $0.9$, so that we can work with 
potentials less or more inclinated, and the $i_b(x)$ distributions used are homogeneous or inhomogeneous (Eq.~(\ref{BiasCurrent})). 
The washboard slope is connected to the heights of the potential barriers seen by the phase elements. 
Reducing the $i_0$ value, the barriers intensity is enhanced and the MST values tend to increase. We search evidences of
nonmonotonic behaviour varying first the values of $L$, $\gamma$ and $\omega$, then the statistics of the noise sources. 
Moreover, we try to find connections between the MST behaviors and JJ soliton dynamics. The amplitude of the oscillating 
driving signal is set to $A=0.7$, to obtain at certain times (see Eq.~(\ref{DrivingCurrent})) $i_b(x,t)>1$ (absence of metastable states) 
and, at least with one of the $i_0$ values used, $i_b(x,t)<0$ (positive slope).
\begin{figure*}[htbp!!]
\centering
\includegraphics[width=181mm]{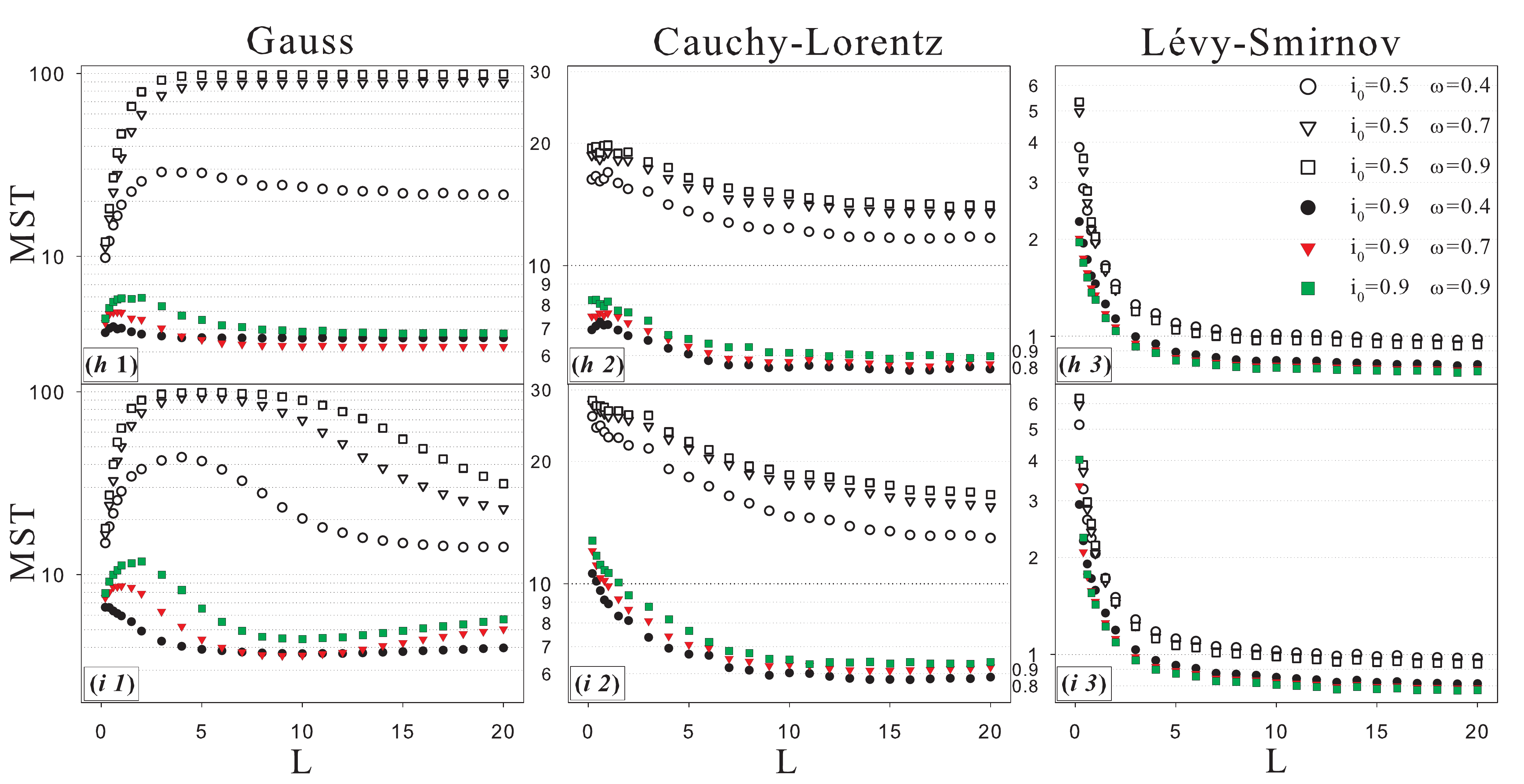}
\caption{(Color online) MST $\tau$ \textit{versus L} for different current distributions along the junction: \emph{homogeneous} $i_b(x)$ and
noise sources with Gaussian (panel (\emph{h 1})), Cauchy-Lorentz (panel (\emph{h} 2)) and L\'evy-Smirnov (panel (\emph{h} 3))
statistics; \emph{inhomogeneous} $i_b(x)$ and noise sources with Gaussian (panel (\emph{i 1})), Cauchy-Lorentz (panel (\emph{i} 2))
and L\'evy-Smirnov (panel (\emph{i} 3)) statistics. In all graphs the other parameters are: $i_0=\left \{0.5 \textup{(empty symbols)},
0.9 \textup{(full symbols)}\right \}$, $\omega=\left \{0.4 \textup{(circles)}, 0.7 \textup{(triangles)}, 0.9 \textup{(squares)}\right \}$ and 
$\gamma=0.2$. The legend in panel (\emph{h} 3) refers to all pictures.} 
\label{fig5+6}.
\end{figure*}
In this section we neglect the thermal fluctuations of the current density \textit{$i_T(x,t)$} with respect to the non-Gaussian 
(L\'evy) noise source \textit{$i_{nG}(x,t)$} in Eqs.~(\ref{sine_gordon}), (\ref{fluct.current}), because we consider very low 
temperatures, around the \textit{crossover} temperature.

\subsection{MST vs JJ length $L$}
\label{MSTvsL}\vskip-0.2cm 

We begin to study the MST values varying the JJ length $L$ in the range $[0,20]$. The results are shown in Fig.~\ref{fig5+6}, 
emphasizing the three different noise sources used, G (panels (\emph{h}1) and (\emph{i}1)), CL (panels (\emph{h}2)
and (\emph{i}2)) and LS (panels (\emph{h}3) and (\emph{i}3)).

\begin{figure*}[htbp!!]
\centering
\includegraphics[width=181mm]{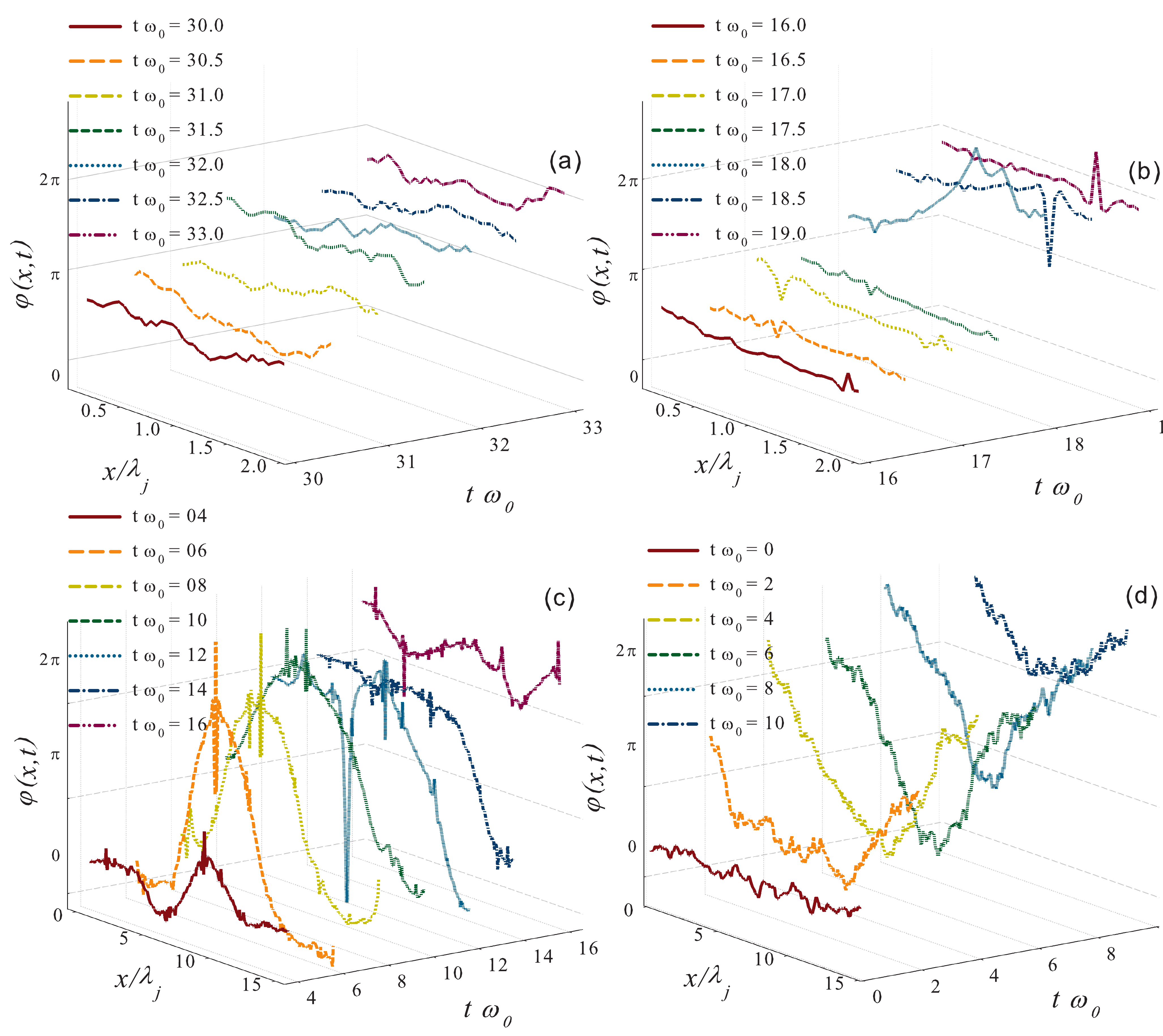}
\caption{(Color online) String dynamics during the switching towards the resistive state: for a JJ of length $L=2$, with homogeneous bias
current distribution and G noise source (panel (a)), inhomogeneous bias current distribution and CL noise source (panel (b)); for a JJ
of length $L=15$, with homogeneous bias current distribution and CL noise source (panel (c)), inhomogeneous bias current distribution
and G noise source (panel (d)). All graphs were obtained for $\omega=0.9$ and $\gamma=0.2$. The curves in panels (b) and (c) show
the characteristic L\'evy flights of the CL statistic.} 
\label{fig7}
\end{figure*}

The panels (\emph{h}1), (\emph{h}2) and (\emph{h}3) contain the results for homogeneous bias current density, while the panels
(\emph{i}1), (\emph{i}2) and (\emph{i}3) contain the results for inhomogeneous bias current density. In each panel, we note that the
MST values for $i_0=0.5$ are greater than those for $i_0=0.9$. This is due to the reduced height of the right potential barrier due to
the increased slope, i.e. $i_0$ value, of the washboard. Specifically the expression for the left (or right) potential barrier 
height $\Delta U^+$(or $\Delta U^-$) is
\begin{eqnarray}
\Delta U^{\pm}(x,t) = &2& \sqrt{1-i_b^2(x,t)}+ \nonumber \\
&+& i_b(x,t)[2\arcsin (i_b(x,t))\pm \pi].
 \label{DeltaU}
\end{eqnarray}
We start analizing the results obtained in the presence of a Gaussian noise source with $i_0=0.5$ and $i_b(x)$ homogeneous 
(empty symbols in the panel (\emph{h}1)). In this panel of Fig.~\ref{fig5+6} it is evident the presence of two different dynamical regimes in 
each of these curves. An initial monotonic increasing behavior is followed by a constant MST plateau. This underlines the presence of two 
different mechanisms, governing the time evolution of the phase, which clearly appear in the soliton dynamics shown in 
Fig.~\ref{fig7}. This picture displays four different phase dynamics during the passage towards the resistive state, i.e. when 
the phase $\varphi$ approximately changes of $2\pi$. The cells can escape from a potential well all together (panel (a) of 
Fig.~\ref{fig7}) or by the formation of a single kink, or a single antikink, or a kink-antikink (K-A) pair (panel (c) of Fig.~\ref{fig7}). 
If the string is too short, the connection among cells is so strong that the soliton formation is forbidden, the string can move from, 
or remain inside, a potential minimum as a whole. This is evident in panel (a) of Fig.~\ref{fig7}. In this length regime, an 
increase in the number of cells makes more difficult the motion of the whole string during the transition process, causing the 
MST to raise for short lengths. This happens as long as no soliton formation occurs. There is, in fact, a specific 
junction length above which the dynamics is governed by the formation of phase kinks. This length is connected with the soliton 
nucleation, that is the formation of a K-A pair. Following the work of B\"{u}ttiker~\cite{Butt81}, in the soliton nucleation a critical 
nucleus, that is the minimum separation between kink and anti-kink, exists. For junction lenghts greater than this critical value 
it is evident a saturation effect. The MST reaches an almost constant value and the switching events are guided by the solitons, 
which indicates that the dynamics of these events is indipendent of the JJ length. To explain this behaviour we consider that 
inside the string a subdomain structure exists. Each subdomain is composed by an amount of cells of total size approximately 
equal to the critical nucleus. The entire string can be thought as the sum of these subdomains and the overall escape event 
results to be the superimposition of the escape events of each single subdomain, so that the total MST is equal to the individual 
subdomain time evolution. The size of this subdomain approximately corresponds to the length value for which the initial 
monotonic behavior is interrupted. The dimension of the critical nucleus is proportional to $L_c \propto -\log (i_0)$. 
Increasing the $i_0$ value, the critical nucleus decreases and the soliton dynamics can start in correspondence of shorter 
junction lengths, as one can see in panel (\emph{h}1) of Fig.~\ref{fig5+6}, where results obtained for $i_0=0.5$ (empty symbols) 
and $i_0=0.9$ (full symbols) are shown. In particular, we have $L_c \sim 5$ for $i_0 = 0.5$, and $L_c \sim 2$ for $i_0 = 0.9$. 
The curves obtained for $i_0=0.9$ are characterized by a small maximum, which reveals the presence of a weak nonmonotonic 
behavior. Between the initial increasing behavior and the saturation, a portion with negative slope and corresponding reduction 
of the MST is evident. Increasing the slope of the potential, the critical nucleus becomes shorter so that the nucleation is 
allowed also in regime of strong connections among the cells. These two conditions, i.e. anticipated nucleation and intense 
\textquotedbl bind\textquotedbl among cells, determine cooperating effects, which lead to MST reduction before the saturation 
regime is reached.

Panels (\emph{h}2) and (\emph{h}3) of Fig.~\ref{fig5+6}  show MST curves obtained in the presence of CL and LS noise sources. 
These behaviors appear quite different with respect to those obtained using Gaussian noise sources. MST curves are strongly 
affected by L\'evy flights that favour jumps between different potential valleys, and soliton  formation (see panel (c) of Fig.~\ref{fig7}, 
containing rapid and sudden phase variations). Specifically, for CL noise the saturation effect gives rise to a value of MST lower 
than that observed with the Gaussian thermal fluctuations. This is due to the peculiarity of the fat tails of PDF for CL noise. 
Therefore, for \textit{homogeneous} density current (panel \textit{h}2), after the initial transient with an increasing behavior due to 
the increasing length of the junction and therefore of the string, nucleation and intense "bind" among cells speed up 
the escape process and $\tau$ decreases towards the saturation value. For \textit{inhomogeneous} density current 
(panel \textit{i}2), the weak nonmonotonic behavior, found for homogeneous case (see panel (\emph{h}2)), disappears.
This is because the edge portions of the phase string are subject to high values of bias current ($i_b(x) > 1$, see Fig.~\ref{fig1} 
and Eq.~(\ref{BiasCurrent})). As a consequence, all the string is dragged out of the potential well, speeding up the 
escape process. The MST values obtained in the presence of LS noise sources are in general smaller than those obtained 
using noise sources with CL distribution. These differences are related to the intensity of the jumps in these two statistics. 
The saturation effect is also present, but the corresponding value of $\tau$ is very low. This is due to the LS L\'evy flights, 
which push the string very fast out of the superconductive state, giving rise to a monotonic decreasing behavior of 
$\tau$ versus $L$. In other words, LS noise drives the phase string out of the potential well very quickly, due to the 
greater diffusive power of this noise source. It is worth noting that, for $i_0=0.9$, the values obtained using the Cauchy-Lorentz 
statistics are slightly greater than those obtained in the presence of Gaussian thermal fluctuations. This is connected 
with the \emph{limited space displacement}, that rules the CL statistics for short time scale~\cite{Auge10}.
\begin{figure*}[htbp!!]
\centering
\includegraphics[width=181mm]{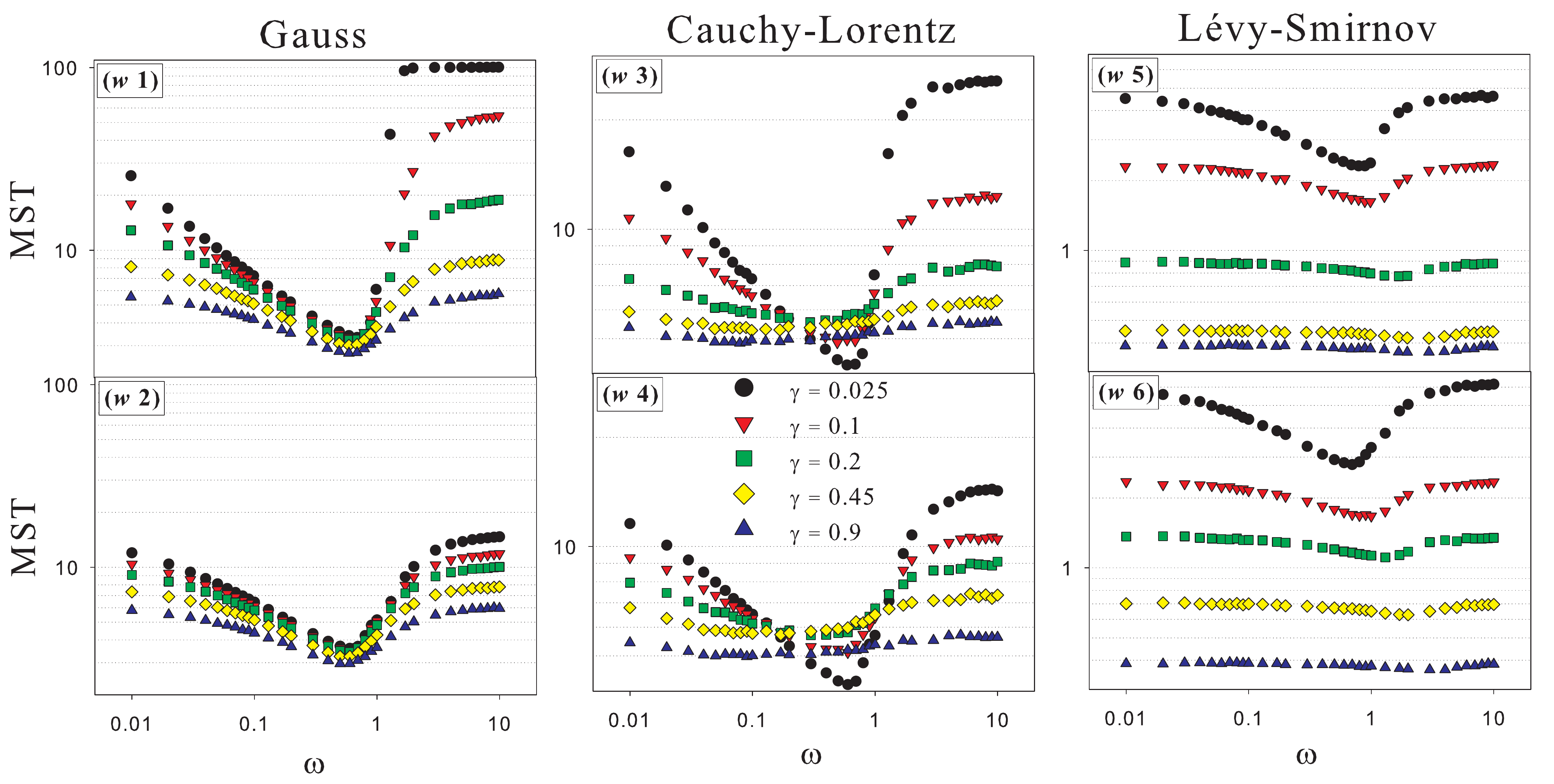}
\caption{(Color online) Log-log plots of MST $\tau$ \emph{versus $\omega$} obtained using: \emph{homogeneous} $i_b(x)$ 
and noise sources G, CL and LS (panels (\emph{w} 1), (\emph{w} 3) and (\emph{w} 5) respectively); 
\emph{inhomogeneous} $i_b(x)$ and noise sources G, CL and LS (panels (\emph{w} 2), (\emph{w} 4) and (\emph{w} 6)
respectively). In all graphs the values of the other parameters are: $i_0=0.9$, $L=10$ and 
$\gamma=\left \{0.025, 0.1, 0.2, 0.45, 0.9\right \}$. The legend in panel (\emph{w} 4) refers to all panels.} 
\label{fig8}
\end{figure*}

In panels (\emph{i}1), (\emph{i}2) and (\emph{i}3) of Fig.~\ref{fig5+6}, we show results obtained in the presence of an
inhomogeneous bias current. According to Eq.~(\ref{BiasCurrent}), $i_b(x)$ diverges at the string ends, $x=0$ and $x=L$, 
having a minimum equal to $i_b(L/2)=2/\pi\cdot i_0$ in the string center, $x=L/2$. In a considerable edge portion of 
the string (around $5\%$ and $18\%$ of the total length for $i_0=0.5$ and 0.9, respectively) $i_b(x)>1$, allowing 
the phase elements to roll down along the tilted potential without encountering any resistance. We can consider 
these edge elements as \emph{generators of solitons}. This corresponds to the physical situation in which the 
supercurrent flows between the junction ends and the fluxon formation occurs in these regions of the JJ. 
This kind of dynamics is shown in panel (d) of Fig.~\ref{fig7}, in which the kink starts from the cells located 
in the junction edges. The role of these cells becomes particularly important as the length $L$ increases, 
but is irrelevant for short junctions, in which the connection between cells is still too strong, and the dynamics
is not guided by solitons. This situation is clear in panel (b) of Fig.~\ref{fig7}, although the presence of CL 
statistics causes the appearance of flights. The G curves in panel (\emph{i}1) of Fig.~\ref{fig5+6} show an increasing 
behavior similar to those obtained with homogeneous bias current distribution, even if the values reached are a 
little bit higher. Independently of the value of $L$, about $77\%$ of the cells composing the junction has $i_b(x)<i_0$. 
Therefore, this percentage of cells should overcome potential barriers higher than those corresponding to the 
case of homogeneous bias current $i_b(x)$. This determines, in the absence of soliton formation, an increase 
of the escape time. Moreover, a nonmonotonic behavior is observed. After reaching the maximum, the MST 
curves decrease due to the action of the junction edges, which behave as generators of solitons. This effect 
accelerates the escape process, becoming more important as the value of $L$ increases (see Fig.~\ref{fig1}).
For $i_0 = 0.9$, the time average of the barrier height is lower than in the case with $i_0 = 0.5$ and the 
switching process is faster.

The CL and LS results presented in panels (\emph{i}2) and (\emph{i}3) of Fig.~\ref{fig5+6} do not show remarkable 
differences with respect to those obtained with homogeneous current distribution, except for an enhancement 
in the MST for very short junction. The physical reason of this behaviour is the same as that discussed for the Gaussian case.

The curves in panels (b) and (c) of Fig.~(\ref{fig7}), obtained using a CL noise source, show peaks associated with the
generations of the L\'{e}vy flights. As previously discussed, these noise induced fluctuations influence the switching events and the
soliton formation. These graphs also clearly display the creation of another "structure", known as \emph{breather} (see panel (b) for
$t\omega _{0}=\left \{ 18.5 , 19 \right \}$ and $x/\lambda_{j}\approx 1.5$, and panel (c)). This is a well-known localized solution of the
SG equation consisting of a soliton-antisoliton pair and oscillating with an internal "breathing" frequency. The curves
obtained by using non-Gaussian noise sources exhibit this kind of nonlinear "structures" (panel (b) and (c) of Fig.~(\ref{fig7})).

\subsection{MST vs driving frequency $\omega$}

\begin{figure*}
\centering
\includegraphics[width=171mm]{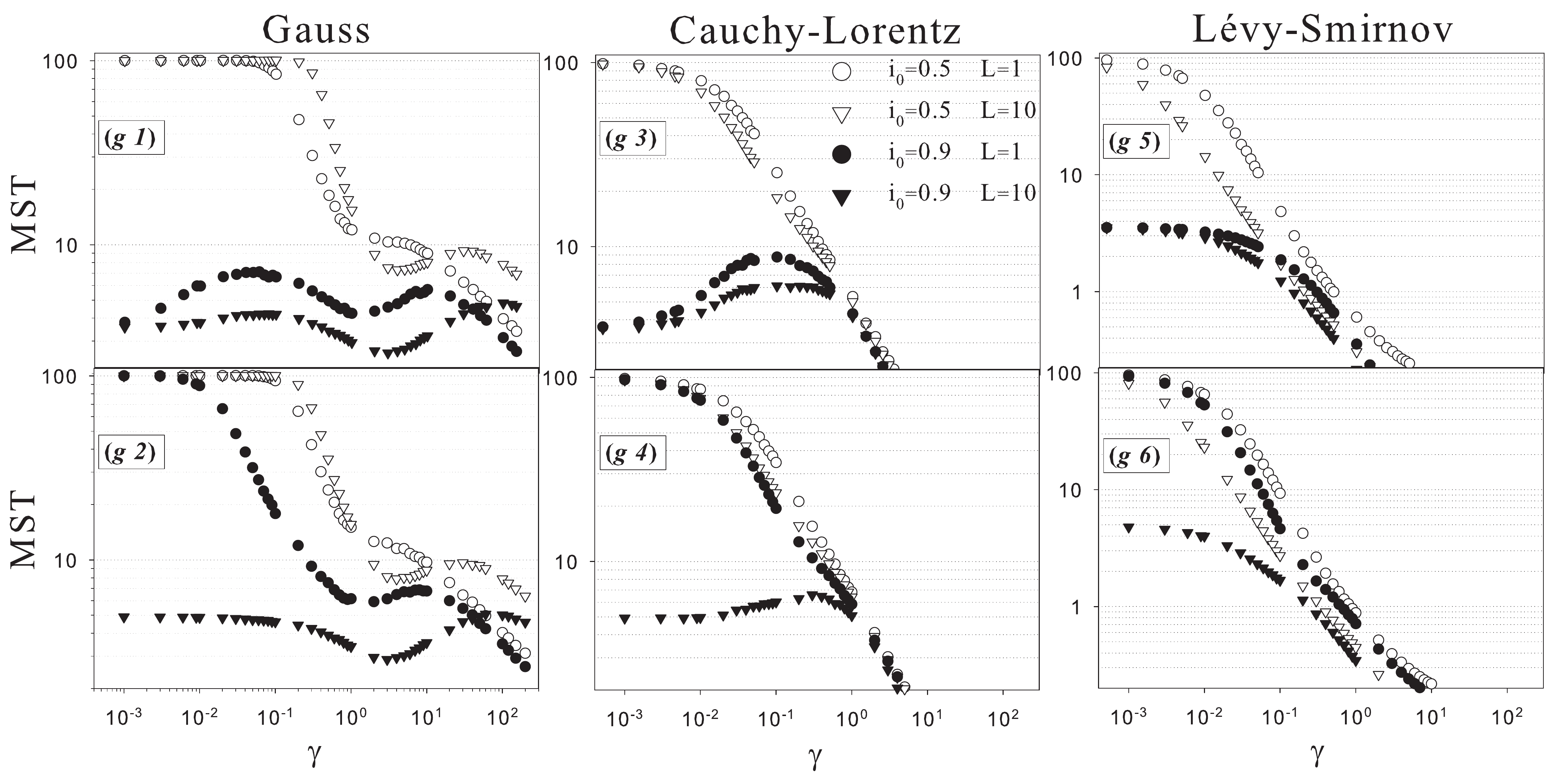}
\caption{(Color online) Log-log plots of MST $\tau$ \emph{versus $\gamma$} obtained using: \emph{homogeneous} $i_b(x)$ 
and noise sources G, CL and LS (panels (\emph{g} 1), (\emph{g} 3) and (\emph{g} 5) respectively); 
\emph{inhomogeneous} $i_b(x)$  and noise sources G, CL and LS (panels (\emph{g} 2), (\emph{g} 4) and (\emph{g} 6)
respectively). In all graphs the values of other parameters are: $i_0=\left \{0.5, 0.9 \right \}$, $\omega=0.9$ and $L=\left \{1,
10\right \}$. The legend in panel (\emph{g} 3) refers to all panels.} 
\label{fig9}
\end{figure*}

In this section we analyze the MST behaviour, setting the bias current at $i_0=0.9$, and varying both the frequency 
$\omega$ of the driving signal (within the interval $[0.01,10]$) and the noise intensity $\gamma$. The values of MST obtained 
are shown in Fig.~\ref{fig8}. Specifically, results obtained in the presence of G, CL and LS noise sources are shown in the
upper panels, (\emph{w} 1), (\emph{w} 3) and (\emph{w} 5) respectively, for homogeneous bias current distribution, 
and in the lower panels (\emph{w} 2), (\emph{w} 4) and (\emph{w} 6) respectively, for inhomogeneous bias current distribution. 
Each panel contains five curves, obtained for the values of $\gamma$ displayed in the legend. This analysis was performed 
working with a junction of length $L=10$, that is a string with a suitable length, which allows to onset the phenomenon of soliton 
formation. All graphs show clearly the presence of \emph{resonant activation} 
(RA)~\cite{Doe92,Man00,Dub04,Man98,Pec94,Mar96,Dyb09,Miy10,Has11,Fias11}, or \textit{stochastic resonance activation}, 
a noise induced phenomenon, whose signature is the appearance of a minimum in the curve of MST \emph{vs} $\omega$.
This minimum tends to vanish for CL and LS distributions when the noise intensities are greater than the time average of the
potential barrier ($\overline{\Delta U}_{i_0=0.9} \simeq 0.4$, see Eq.~(\ref{DeltaU})). It is worthwhile to note that the nonmonotonic 
behavior of $\tau$ versus the CL noise intensity around the minimum, observed in panels (\emph{w} 3) and (\emph{w} 4) of Fig.~\ref{fig8}, 
is related to that shown in panels (\emph{g} 3) and (\emph{g} 4) of Fig.~\ref{fig9}.
The RA is a phenomenon robust enough to be observed also in the presence of L\'{e}vy noise sources~\cite{Auge10}.
Particle escape from a potential well is driven when the potential barrier oscillates on a time-scale characteristic of the particle
escape itself. Since the resonant frequency is close to the inverse of the average escape time at the minimum, which is
the mean escape time over the potential barrier in the lower configuration, \textit{stochastic resonant 
activation} occurs~\cite{Add12,Pan09}, which is a phenomenon different from the \textit{dynamic resonant activation}. This effect, in fact, 
appears when the driving frequency matches the natural frequency of the system, that is the plasma frequency~\cite{Dev84,Dev85,Mar87}. 
Finally, we note that the contemporaneous presence of RA and NES phenomena in the behavior of $\tau$ as a function of the 
driving frequency, in underdamped JJ, has been observed, finding that the MST can be enhanced or lowered by using 
different initial conditions~\cite{Sun07}. 

The G data in panels (\emph{w} 1) and (\emph{w} 2) of Fig.~\ref{fig8} present this minimum for a frequency value 
($\omega_{RA} \sim 0.6$) which varies little with the noise intensity $\gamma$. The only evident effect, 
switching to an inhomogeneous bias current, is a general reduction of the MST. The curves with CL noise 
present a clear minimum, shifted towards higher values of $\omega$, with respect to that of the Gaussian case. 
This minimum tends to disappear increasing the noise intensity. This is due to the influence of L\'evy flights
which, for strong noise intensities, drive the escape processes. As found in the presence of Gaussian noise, 
also in the case of CL statistics, using inhomogeneous $i_b(x)$ causes a general lowering in the MST values. 
We can note that for a weak noise signal, the Cauchy-Lorentz distributions are higher than the Gaussian ones: 
for low values of $\gamma$ the jumps are not relevant, and the limited space displacement gives short phase 
fluctuations, making more difficult to escape from the potential wells. The MST calculated using LS sources are 
also governed only by the noise and present quite small values. Therefore, the RA effect is found only in the curve 
obtained for a very weak noise intensity.

By increasing the driving frequency, at low noise intensities, a trapping phenomenon occurs. A threshold frequency 
 $\omega_{thr}$ exists such that for  $\omega > \omega_{thr}$ the phase string is trapped within a region 
 between two successive minima of the potential profile. In other words, the string can not move 
 from the potential well to the next valley during one period $T_0$ of the driving current $A sin(\omega t)$. 
 As a consequence, the MST diverges in the limit $\gamma \rightarrow 0$. The value of the threshold frequency 
 increases with increasing bias current and/or maximal current across the junction~\cite{Gord08,Dub04,Agu01}. 
 We have estimated the threshold values for the following parameter values $i_0 = 0.9$ and $A = 0.7$. 
 Specifically, for Guassian thermal fluctuations $\omega_{thr} \apprge 1.8$, for CL noise 
 source $\omega_{thr} \apprge 2.1$ and for LS noise source $\omega_{thr} \apprge 3$.

\subsection{MST vs noise intensity $\gamma$}

Here we analyze the MST curves calculated varying the noise amplitude in the range $[0.0005,200]$. 
The results are shown in Fig.~\ref{fig9}. Specifically the results in panels (\emph{g} 1), (\emph{g} 3) and (\emph{g} 5)
were obtained, using an homogeneous $i_b(x)$ and G, CL and LS noise sources respectively, while those shown 
in panels (\emph{g} 2), (\emph{g} 4) and (\emph{g} 6), using an inhomogeneous $i_b(x)$ and G, CL and LS noise
sources respectively. This analysis is performed using $\omega=0.9$ and two different values of $L$ and $i_0$, 
i.e. $L=\left \{1, 10\right \}$ and $i_0=\left \{0.5, 0.9 \right \}$. Fixing the values of the system
parameters, for $\gamma\rightarrow 0$ the curves for the three noise sources (G, CL and LS) converge to the same
values, i.e. the deterministic lifetime in the superconducting state, which depend strongly on the bias current. 
When $\gamma \rightarrow 0$ and the potential is not too tilted, trapping phenomena occur and the MST tends 
to $t_{MAX}$. Increasing the noise intensity, the MST curves exhibit an effect of \emph{noise enhanced
stability} (NES)~\cite{Dub04,Man96,Agu01,Spa04,D'Od05,Fia05,Hur06,Spa07,Mank08,Yos08,Fia09,Tra09,Fia10,Li10,Smi10}, 
a noise induced phenomenon consisting in a nonmonotonic behaviour with the appearance of a maximum. 
The stability of metastable states can be enhanced and the average life time of the metastable
state increases nonmonotonically with the noise intensity. The observed nonmonotonic resonance-like
behavior proves to be different from the monotonic one of the Kramers theory and its 
extensions~\cite{Kra40,Mel91,Han90}. This enhancement of stability, first noted by Hirsch et al.~\cite{Hir82}, has 
been observed in different physical and biological systems, and belongs to a highly topical interdisciplinary 
research field, ranging from condensed matter physics to molecular biology and to cancer growth 
dynamics~\cite{Spa07,Spa12}.
\begin{figure*}
\centering
\includegraphics[width=171mm]{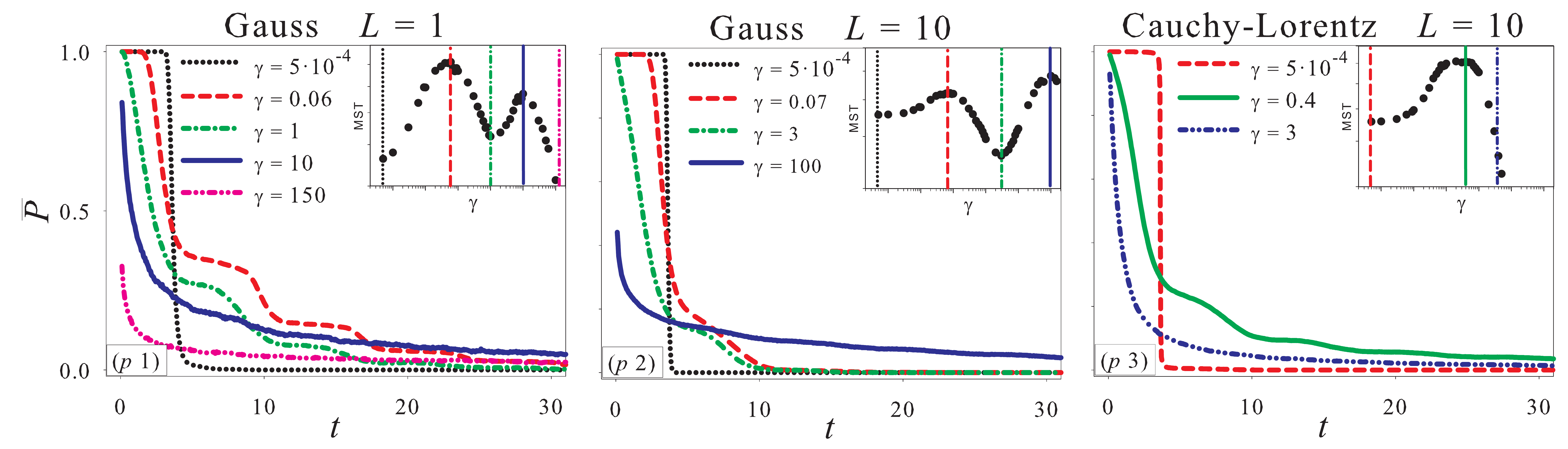}
\caption{(Color online) Time evolution of the probability $P(t)$ in the following conditions: G noise with $L=1$ 
(panel (\emph{p} 1)) and $L=10$ (panel (\emph{p} 2)); CL noise with $L=10$ (panel (\emph{p} 3)). 
The system parameters are $i_0=0.9$ and $\omega=0.9$. Each graph contains curves of $P(t)$ obtained using values 
of $\gamma$ for which a minimum or maximum appears in the \textit{$\tau$ vs $\gamma$} behaviour. The insets reproduce 
the corresponding curves of panels (\emph{g} 1) and (\emph{g} 3) of Fig.~\ref{fig9}.} 
\label{fig10}
\end{figure*}

From Fig.~\ref{fig9}, we note that in the curve obtained using a Gaussian noise source, homogeneous 
current distribution and high washboard inclination, $i_0=0.9$, two maxima are present in correspondence 
of $\gf\cong\left \{0.06, 10\right \}$ for $L=1$ and $\gs \cong \left \{0.07, 100\right \}$ for $L=10$. 
In view of understanding the physical motivations of these NES effect, we calculate the time evolution of 
the probability $P(t)$, as defined in Eq.~(\ref{P_averaged}), during the switching dynamics of the junction. 
We remember that $0\leq P(t)\leq 1$, where the two extreme values indicate the \textup{resistive state} 
($P(t)=0$) and the \textup{superconducting state} ($P(t)=1$).

The time evolution of $P(t)$ was calculated for $i_0=0.9$ and $\omega=0.9$. The results, shown in Fig.~\ref{fig10}, 
were obtained in the following conditions: i) G noise with $L=1$ (panel (\emph{p} 1)) and $L=10$ (panel (\emph{p} 2)); ii) 
CL noise with $L=10$ (panel (\emph{p} 3)). All panels of Fig.~\ref{fig10} contain curves of $P(t)$ calculated setting the noise 
intensity at values for which a maximum or minimum appears in the MST vs $\gamma$ behaviour (see insets). Looking at 
the curves displayed in panel (\emph{p} 1), we note that the dotted curve ($\gamma=0.0005$) represents a deterministic 
switching event. The string after a quick escape does not return inside the first washboard valley. Conversely, the dashed line, 
obtained for $\gamma=0.06$, describes a temporary trapping phenomenon. The contemporaneous presence of the fluctuating 
potential and noise source, inhibits the phase switching and therefore the passage of the junction to the resistive regime. 
Moreover the exit from the first well is not sharp, as in the deterministic case, and $P(t)$ assumes an oscillatory behavior, 
almost in resonance with the periodical motion of the washboard potential. This oscillating behavior of $P(t)$, which is 
related to the temporary trapping of the phase string, tends to disappear as the noise intensity increases.
For $\gamma=10$ (solid line in (\emph{p} 1) of Fig.~\ref{fig10}) another peak (NES effect) in the MST behaviour is 
observed, but no oscillations in $P(t)$ are present. At this value of $\gamma$, the JJ dynamics is totally driven by the 
noise and the NES effect is due to the possibility that the phase string returns into the first valley after a first escape event, 
as indicated by the fat tail of $P(t)$. This behaviour is strictly connected with that found for the MST, whose calculation is 
based on the definition of NLRT. Further increases of $\gamma$ reduce for the phase string the possibility not only of 
returning into the intial well but also of staying for a long time inside it. The results for G noise source and $L=10$, 
displayed in panel (\emph{p} 2) of Fig.~\ref{fig10}, are similar to those obtained for $L=1$. The first hump, corresponding 
to $\gamma=0.07$ (see inset of panel (\emph{p} 2)) is a little bit smaller than that for $L=1$ and $\gamma=0.06$ 
(see inset of panel (\emph{p} 1)), and this is consistent with the previous MST \emph{versus L} analysis. 
Moreover a NES effect for $\gamma=100$ is present (see inset of panel (\emph{p} 2)). We note the difference of one 
order of magnitude in the noise intensity ($\gamma=100$ for $L=10$) respect to the NES phenomenon observed for $L=1$ at 
$\gamma=10$. This difference is due to the greater difficulty for random fluctuations of carrying a string, ten times longer, again 
in the initial potential well.
\begin{figure*}
\centering
\includegraphics[width=171mm]{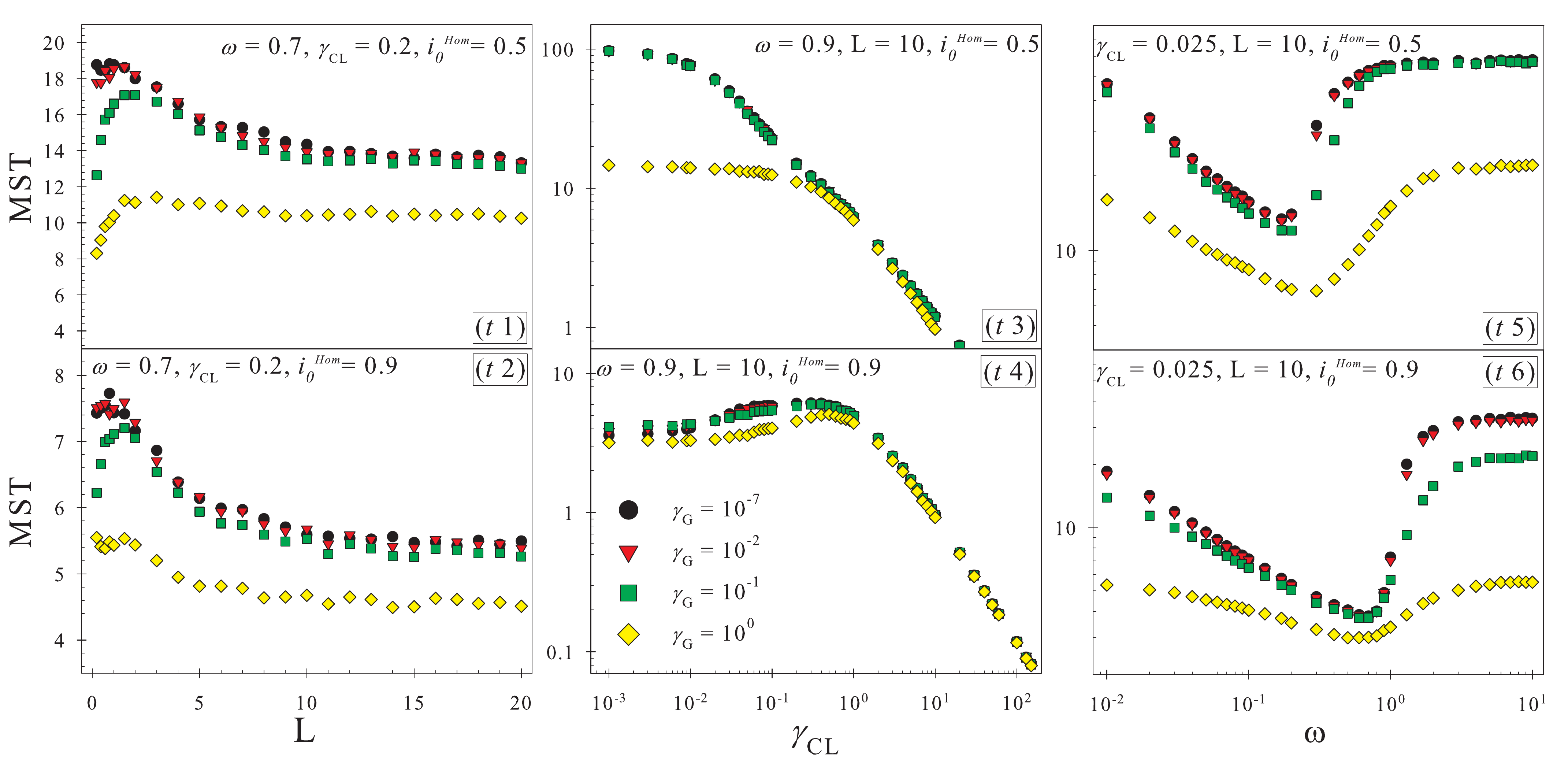}
\caption{(Color online) MST $\tau$ as a function of $L$, $\gamma_{\mathrm{CL}}$, and $\omega$. 
All curves were obtained considering the simultaneous presence of CL and thermal noise sources, using two different values, 
i.e. $i_0=0.5$ (top panels) and $i_0=0.9$ (bottom panels), of the homogenous bias current, and varying the Gaussian noise intensity, 
$\gamma_{\mathrm{G}}$. The legend in panel (\emph{t} 4) refers to all panels.} 
\label{fig11}
\end{figure*}
Panel (\emph{g} 3) of Fig.~\ref{fig9} shows the curves of MST vs $\gamma$ in the presence of CL noise source. Here we note the
absence of the second peak, observed in the previous analysis at higher values of $\gamma$. This discrepancy can be explained noting
that, for low noise intensity, the effect of the CL flights on the overall JJ dynamics is neglegible, and the time evolution should
appear quite similar to those observed with Gaussian noise. Conversely, due to the limited space displacement, to obtain the
same effect (i.e. escape from the first potential well), junctions subject to CL noise should be exposed to noise intensity larger
respect to identical junctions subject to G noise. The peak (maximum of MST) should be therefore shifted towards higher values of
$\gamma$. Increasing the noise intensity, the influence of L\'evy-flights on the total JJ dynamics becomes higher, and the
probability that a second peak appears, similar to that observed in the presence of G noise, tends to vanish. This analisys is confirmed
by the graph shown in panel (\emph{p} 3) of Fig.~\ref{fig10}. Conversely, LS flights are too intense to allow the formation of NES peaks (see
panel (\emph{g} 5) of Fig.~\ref{fig9}). Finally, we note that the curves obtained with inhomogeneous $i_b(x)$ do not present any
differences, except those for $i_0=0.9$ and $L=1$ (full circles), that show very high values of MST with respect to the case of
homogenous current distribution. This indicates again a trapping phenomenon that occurs when a short junction undergoes very weak
noise intensities ($\gamma \rightarrow 0$). In this case, the parts of the junction generating solitons do not affect the string dynamics. 
In fact, since $i_b(x)<i_0$ for 77$\%$ of the total length, a large percentage of the string remains confined in a potential well 
deeper than that of the analogous homogeneous case, thus determining the trapping effect.

Moreover, all the curves of MST vs $\gamma$ for CL and LS noise sources coalesce together at high noise intensities. The MST has a 
power-law dependence on the noise intensity according to the expression
\begin{equation}
\tau \simeq \frac{C(\alpha)}{\gamma^{\mu(\alpha)}}
\label{power-law}
\end{equation}
where the prefactor C and the exponent $\mu$ depend on the L\'evy index $\alpha$~\cite{Dub08}.
From Fig.~\ref{fig9} we have $\mu(\alpha) \sim 0.9$ for CL noise and $\mu(\alpha) \sim 1.2$ for LS noise, which are in 
agreement with the exponent $\mu(\alpha) \approx 1$ for $0 < \alpha < 2$, calculated for barrier crossing 
in bistable and metastable potential profiles~\cite{Che05,Che07}.

\section{Simultaneous presence of L\'evy noise and thermal fluctuations}

In this section we analyze the presence of both thermal and L\'evy noise sources. Therefore, in Eqs.~(\ref{sine_gordon}) 
and~(\ref{fluct.current}) both contributions of Gaussian thermal fluctuating current density \textit{$i_T(x,t)$} and 
non-Gaussian L\'evy noise current density \textit{$i_{nG}(x,t)$} are considered. 
The L\'evy contribution is restricted to a Cauchy-Lorentz term. The noise intensities are indicated by 
$\gamma_{\mathrm{G}}$ (Gaussian), ranging within the interval $[10^{-7},1]$, and $\gamma_{\mathrm{CL}}$ (Cauchy-Lorentz). 
Noise induced phenomena previously observed, when the CL noise source only is present, show now some differences. The values of the
system parameters are chosen in such a way to highlight these changes. Fig.~\ref{fig11} contains a collection of MST curves
obtained varying the junction lenght $L$ (panels (\emph{t} 1) and (\emph{t} 2)), CL noise intensity $\gamma_{\mathrm{CL}}$ (panels
(\emph{t} 3) and (\emph{t} 4)), and frequency of the oscillating bias current $\omega$ (panels (\emph{t} 5) and (\emph{t} 6)). Top
and bottom panels show data calculated using $i_0=0.5$ and $i_0=0.9$, respectively. An overall reduction of the MST values is
observed by increasing the intensity of thermal fluctuations, by speeding up the switching process between the superconductive 
and the resistive state. The simultaneous presence of thermal fluctuations and a L\'evy noise source produces an increase of 
the overall intensity "felt" by the string phase. In all panels clear modifications of the nonmonotonic behavior are present, 
becoming more pronounced as the Gaussian thermal noise intensity increases, especially for $\gamma_{\mathrm{G}}>10^{-1}$. 

The analysis of MST vs $L$ suggests that the soliton dynamics is modified only when the intensity of thermal fluctuations
are greater than those of the CL noise, that is $\gamma_{\mathrm{G}} > \gamma_{\mathrm{CL}}$, 
conversely the curves for $\gamma_{\mathrm{G}} < \gamma _{\mathrm{CL}}$ overlap all together 
($\gamma_{\mathrm{G}} \leq 10^{-1} $). The curves of the panels (\emph{t} 1) and (\emph{t} 2) maintain the structure 
already shown in panel  (\emph{h} 2) of Fig.~\ref{fig5+6} (see Sec.~\textbf{V\,A}), that is a nonmonotonic 
behavior with a maximum and a saturation plateau. The saturation value of $\tau$ decreases, of course, with the increase
of the intensity of thermal fluctuations. 

Looking at the graphs of MST vs $\gamma_{\mathrm{CL}}$ (panel (\emph{t} 3)), 
trapping phenomena are observed when $\gamma_{\mathrm{CL}} \rightarrow 0$ and $\gamma_{\mathrm{G}} \rightarrow 0$. 
For $\gamma_{\mathrm{G}} \ge 1$, that is when the Gaussian thermal noise intensity is comparable with the 
time average of the potential barrier height ($\overline{\Delta U}_{i_0=0.5} \simeq 1$, see Eq.~(\ref{DeltaU})), trapping events disappear 
and thermally activated processes drive the switching events. For $i_0=0.9$ (panel (\emph{t} 4)) all the curves 
show a nonmonotonic behavior, which is the signature of the NES effect. Low thermal noise intensities do not affect 
the behavior of the NES curve, with respect to the case of absence of thermal noise, till their value is lower 
than $\gamma_G \simeq 0.2$. This is the value of the CL noise intensity corresponding to the maximum of 
$\tau$ versus $\gamma_{\mathrm{CL}}$, $\gamma^{Max}_{\mathrm{CL}} \simeq 0.2$ (see panel (\emph{g} 3) of Fig.~\ref{fig9}). 
In other words, thermal fluctuations affect the behavior of NES curve for $\gamma_G \gtrsim \gamma^{Max}_{\mathrm{CL}}$. 
The maximum of the curve decreases and it is shifted towards higher CL noise intensities, because of the larger spatial 
region of the potential profile spanned by the phase string before reaching the boundaries [$-\pi,\pi$].

For CL noise intensities $\gamma_{CL} \gtrsim 1$, all the curves of MST vs $\gamma_{CL}$ 
(see panels (\emph{t} 3) and (\emph{t} 4)) coalesce together with a power-law behavior given by Eq.~(\ref{power-law}), 
with $\mu(\alpha) \sim 0.9$. When the structure of the potential profile becomes irrelevant for the dynamics of the
phase string, that is when the noise intensity $\gamma_{CL}$ is greater than the time average of the potential 
barrier heights ($\overline{\Delta U}_{i_0=0.5} \simeq 1$ and $\overline{\Delta U}_{i_0=0.9} \simeq 0.4$), 
the MST has a power-law dependence on the noise intensity.

The curves of MST as a function of $\omega$ in panels (\emph{t} 5) and (\emph{t} 6) of Fig.~\ref{fig11} reproduce 
the typical RA behavior (see panels  (\emph{w} 3) and (\emph{w} 4) of Fig.~\ref{fig8}). Again, all the curves of MST 
are lowered for increasing thermal fluctuation intensities. Specifically, for $i_0 = 0.5$ (panel (\emph{t} 5)),  the minimum 
of the curve decreases and it is shifted towards higher values of the driving frequency. The resonant rate escape, 
that is the resonant frequency at the minimum, increases by increasing the overall noise intensity, being fixed the 
height of the average potential barrier ($\overline{\Delta U}_{i_0=0.5} \simeq 1$). For $i_0 = 0.9$ (panel (\emph{t} 6)), 
there is not any potential barrier for about half period of the external driving force, and therefore the switching process 
is accelerated, and the position of the minimum is slightly affected by thermal fluctuations.

\section{Conclusions}

We have investigated the influence of both thermal fluctuations and external non-Gaussian noise sources on 
the temporal characteristics of long-overlap JJs. We studied how random fluctuations with different 
$\alpha$-stable (or L\'{e}vy) distributions affect the superconducting lifetime of long current-biased Josephson 
junctions. The study was performed within the framework of the sine-Gordon equation. Specifically we analyzed 
the mean switching time (MST) of the phase difference across the junction, from a minimum of the tilted washboard potential, 
as a function of different parameters of the system and external random and periodical driving signals. We found 
nonmonotonic behaviors of the superconducting lifetime $\tau$ as a function of noise intensity $\gamma$, 
driving frequency $\omega$ and junction length $L$. 

In particular, in the behaviour of the MST, we observed noise induced phenomena such as \emph{stochastic resonant activation} 
and \emph{noise enhanced stability}, with different characteristics depending on both the bias current distribution 
along the junction and the length of the superconducting device. Moreover, temporary trapping of the phase string in the 
metastable state with Gaussian thermal and  CL noises gives rise to an oscillating behavior of the time evolution of the 
probability $P(t)$. The analysis of the MST as a function of the junction length revealed that the \emph{soliton dynamics} 
plays a crucial role in the switching dynamics from the superconducting to resistive state. In more detail, we studied 
the relationship between creation and propagation of solitons and different features of the mean switching time. 
This analysis has demonstrated the existence of two different dynamical regimes. One, occurring for short junction, is characterized 
by the movement of the phase string as a whole. The other one, occurring for junction whose size exceeds a critical length, 
in which the kink (or antikink) creation is allowed.

Moreover, for high values of the bias current, there is a length in which the two regimes take place 
simultaneously. Finally we found that, choosing an inhomogeneous distribution of the bias current along the junction, 
the cells located at the junction edges behave as generators of solitons. In these conditions the escape from the metastable 
superconducting state is strongly affected by the soliton dynamics. The analysis of the contemporaneous presence of 
Cauchy-Lorentz and thermal noise sources gives rise to modifications in the soliton dynamics and noise induced effects 
observed in the transient dynamics of JJs in the presence of non-Gaussian, L\'evy type noise sources. Moreover oscillating 
pairs of soliton-antisoliton (\textit{breathers}) induced by the noise have been observed.

Our findings, which are important to understand the physics of fluctuations in long-overlap Josephson junctions 
to improve the performance of these devices, could help to shed new light on the general context of the nonequilibrium 
statistical mechanics. In fact, JJs are good candidates for probing relevant physics issues in metastable systems~\cite{Sun07}. 
Moreover, the mean switching time from one of the metastable states of the potential profile encodes information on the 
non-Gaussian background noise. Therefore, the statistical analysis of the switching times of JJs can be used to analyze 
weak signals in the presence of an unknown non-Gaussian background noise.

\section*{Acknowledgements}

Authors acknowledge the financial support of Ministry of Education, University, and Research of Italian
Government (MIUR).

\end{document}